\documentstyle[12pt]{article}
\title{\begin{flushright}
{\normalsize NUC-MINN-95/19-T\\
July 1995 \\}
\vspace*{0.25in}
\end{flushright}
{\bf INHOMOGENEOUS NUCLEATION OF QUARK--GLUON PLASMA IN HIGH
ENERGY NUCLEAR COLLISIONS}}
\author{{\bf Joseph I. Kapusta} and {\bf Axel P. Vischer$^{\dagger}$}\\
  {\it School of Physics and Astronomy}\\
   {\it University of Minnesota}\\ {\it Minneapolis, MN 55455}}

\date{}

\begin{document}

\maketitle

\begin{center}
Abstract
\end{center}

\noindent
We estimate the probability that a hard nucleon-nucleon collision
is able to nucleate a seed of quark--gluon plasma in the surrounding
hot and dense hadronic matter formed during a central collision of two
large nuclei at AGS energies.  The probability of producing at least
one such seed is on the order of 1-100\%.  We investigate the
influence of quark--gluon plasma formation on the observed multiplicity
distribution and find that it may lead to noticeable structure in the
form of a bump or shoulder.

\vspace*{1.0in}

\flushbottom
\noindent
-----------------------------------------------------------------------\\
$\dagger$ Address after 1st of October:
Niels Bohr Institute, DK-2100, Copenhagen \O, Denmark.
\vfill \eject

\section{Introduction}

One of the mysteries of heavy ion physics at Brookhaven National
Laboratory's AGS is: {\it If hadronic cascade event simulators like
RQMD \cite{rqmd} and ARC \cite{arc} produce energy densities
approaching 2 GeV/fm$^3$, yet agree with experiment, where is
the quark--gluon plasma?}  After all, numerous estimates of the
onset of quark--gluon plasma agree that it should occur at about
that energy density, and if there is a first order phase transition,
then the onset of the mixed phase would occur at an even lower density.
One possibility is that no phase transition occurs even at these
high densities, but it is difficult to understand how composite
objects like hadrons can overlap so strongly in position space
without the matter undergoing some qualitative change in character.
A second possibility is that the distribution of observed hadrons
in the final state is insensitive to the dynamics of the matter
when it is most hot and dense.  (Unfortunately there are no
measurements of direct photons or dileptons at the AGS which might
probe this stage of the collision.)  There is some evidence for
this which comes from artificially modifying hadronic cross sections
at high density \cite{pang}.  It may be understood by recognizing
that once a system reaches local thermal equilibrium it is basically
irrelevant how it got there.

Recently we proposed a third possibility \cite{us}: {\it Most collisions
at AGS energies produce superheated hadronic matter and are describable
with hadronic cascade simulators, but in rare events a droplet of
quark--gluon plasma is nucleated which converts most of the matter to
plasma.}  We estimated the probability of
this to occur, using homogeneous nucleation theory, to be on the
order of once every 100 to 1000 central collisions of large nuclei.
Our estimate was based on the probability that thermal
fluctuations in a homogeneous superheated hadronic gas would produce
a plasma droplet, and that this droplet was large enough to overcome
its surface free energy to grow.  In this paper we consider another
source of plasma droplet production which is essentially one of
nonthermal origin. Specifically, we estimate the probability
that a collision occurs between two highly energetic incoming nucleons,
one from the projectile and one from the target,
that this collision would have produced many pions if it had occurred
in vacuum, but because it occurs in the hot and dense medium
its collision products are quark and gluon fields which make a small
droplet of plasma.  Although there is a large uncertainty in our
estimates, we find that this inhomogeneous nucleation of plasma may be
more probable than homogeneous nucleation by one to two orders of magnitude.

In this paper we also consider the problem of observation of the
effects of nucleation of plasma in rare events.  We are guided by
observations of multiplicity distributions in $p\bar{p}$ collisions
at the CERN and Fermilab colliders.  In those distribution, one
sees a shoulder developing at high multiplicity at an energy of
540 GeV, which turns into a noticeable bump at higher energies.
The real cause of this structure is not known, but may be due to
minijet production.  If plasma is nucleated in some fraction of
central nucleus--nucleus collisions at the AGS, a similar structure
may develop.

\section{Kinetic Model of Hard Nucleon-Nucleon Collisions}

In this section we develop a simple kinetic model which allows us
to estimate the number of high energy nucleon-nucleon scatterings
occurring in the high density medium formed during a collision
between heavy nuclei.  These scatterings occur when a projectile
nucleon penetrates the hot and dense matter to collide with
a target nucleon which has also penetrated the hot and dense matter.
The energy loss of the colliding nucleons must be taken into
account to obtain a reasonable estimate of the energy available
for meson production in the nucleon-nucleon collision.

To first approximation we can visualize the initial stage of a
heavy ion collision at the AGS in the nucleus-nucleus center-of-
momentum frame as two colliding Lorentz contracted disks.
See Figure 1.  At time $t = 0$ they touch; subsequently they
interpenetrate, forming hot and dense matter in the region of overlap.
During this stage, additional matter streams into the hot zone
even as this zone is expanding along the beam axis.  The nucleons
streaming in undergo scatterings with the hot matter already present,
degrading their longitudinal momentum and producing baryonic isobars
and/or mesons.  Finally, at time $t_0 = L/2v\gamma$, all the
cold nuclear matter has streamed into the region of overlap,
and expansion and cooling begins.  Here, $L$ is the nuclear thickness,
$v$ is the velocity in the center-of-momentum frame, and $\gamma$
is the associated Lorentz contraction factor.  This is a very
simplified picture of the early stage of the collision, but it
seems to semi-quantitatively represent the outcome of both
the ARC and RQMD simulations \cite{rqmd,arc,us}.

We are interested in the possibility that an incoming projectile
nucleon suffers little or no energy loss during its passage
to the longitudinal point $z$ inside the hot and dense zone
where it encounters a target nucleon which also has suffered
little or no energy loss.  The energy available in the ensuing
nucleon-nucleon collision, $\sqrt{s}$, can go into meson production.
Suppose that a large number of pions would be produced if the collision had
happened in free space.  Clearly, the outgoing quark and gluon fields
cannot be represented as asymptotic pion and nucleon states immediately.
The fields must expand and become dilute enough to be called real
hadrons.  If this collision occurs in a high energy density medium,
the outgoing quark-gluon fields will encounter other hadrons before
they can hadronize.  It is reasonable to suppose that this ``star
burst" will actually be a seed for quark-gluon plasma formation
if the surrounding matter is superheated hadronic matter.  We need
a semi-quantitative model of this physics.

A fundamental result from kinetic theory is that the number of
scattering processes of the type 1 + 2 $\rightarrow X$ is given by
\begin{equation}
N_{1+2 \rightarrow X}
= \int dt \int d^3 x \int \frac{d^3 p_1}{(2 \pi )^3} \, f_1 ({\bf x},
{\bf p}_1,t) \int \frac{d^3 p_2}{(2 \pi )^3} \, f_2 ({\bf x},{\bf p}_2,t)
\, v_{12} \, \sigma_{1+2 \rightarrow X}(s_{12}) \, .
\end{equation}
Here $v_{12}$ is a relative velocity,
\begin{equation}
v_{12} = \frac{\sqrt{(p_1 \cdot p_2)^2 -m_N^4}}{E_1 \, E_2} \, ,
\end{equation}
where $p_i$ denotes the four-momentum of nucleon $i$ and $E_i =\sqrt{
{\bf p}_i^2 + m_N^2}$ its energy.  The $f_i$ are phase space
densities normalized such that the total number of nucleons of type
$i$ is
\begin{equation}
N_i^{\rm tot} = \int \frac{d^3x d^3p}{(2\pi)^3} f_i({\bf x,p},t) \, .
\end{equation}
A differential distribution in the variable $Y$ is obtained by replacing
$\sigma$ with $d\sigma/dY$.

For our purpose it is reasonable to represent the colliding
nuclei as cylinders with radius $R$ and thickness $L$.  All the
action is along the beam axis.  We assume that the phase space
distributions are independent of transverse coordinates $x$ and
$y$ and of transverse momentum.  Integrating over the cross sectional
area of the nuclei, and counting only those collisions that occur
within the hot zone, yields
\begin{equation}
N_{1+2 \rightarrow X} = \pi R^2 \int_0^{t_0} dt \int_{-vt}^{vt} dz
\int \frac{dp_{1z} \, dp_{2z}}{(2 \pi)^2}
f_1 (z,p_{1z},t) \, f_2 (z,p_{2z},t) \, v_{12}
\, \sigma_{1+2 \rightarrow X}(s_{12}) \, .
\end{equation}
Here there is a change in notation: $f_i(z,p_{iz},t)/2\pi$ is the
probability per unit volume to find a nucleon $i$ with longitudinal
momentum $p_{iz}$ at longitudinal position $z$ at time $t$.
The integration limits on $z$ ensure that the collisions under
consideration really occur in the hot zone; see Figure 1.
The integration limits on $t$ mean that we only count those
collisions which occur before the system begins its cooling stage.
The depth in the hot zone to which nucleon 1 has penetrated is
$d_1 = (vt+z)/2$, and the depth to which nucleon 2 has penetrated
is $d_2 = (vt-z)/2$.  We neglect the decrease in velocity of the
nucleons as they travel through the hot zone.  This is an acceptable
approximation because in the end we are interested only in those
nucleons which suffer a small energy loss in traversing the hot
matter.

We construct the phase space distribution as follows:
\begin{verse}
$H(x,N)$ = probability that the nucleon has momentum fraction
$x$ after making N collisions;\\
$S(N,d)$ = probability that the nucleon has made $N$ collisions
after penetrating to a depth $d$;\\
$\sum_{N=0}^{\infty} H(x,N) S(N,d)$ = probability that the nucleon
has momentum fraction $x$ after penetrating to a depth $d$.
\end{verse}
The distribution functions are normalized to unity.
\begin{eqnarray}
\int_{0}^{1} \frac{dx}{x} \, H(x,N) &=& 1 \\
\sum_{N = 0}^{\infty} \, S(N,d) &=& 1
\end{eqnarray}
The phase space density of nucleon $i$ is then taken to be
\begin{equation}
\frac{dp_{zi}}{2\,\pi} \, f_i (z,p_{iz},t) =
\gamma \, n_0 \, \frac{dx_i}{x_i} \sum_{N_i = 0}^{\infty}
\, H(x_i,N_i) \, S(N_i,d_i) \, ,
\end{equation}
where $n_0$ is the average baryon density in a nucleus, about 0.145
nucleons/fm$^3$.  As a check, we can compute the number of nucleons
which have entered the hot zone as a function of time.
\begin{equation}
N_i^{\rm part}(t) = \int \frac{d^3x dp_{iz}}{2\pi} f(z,p_{1z},t) \, \Theta
(d_i)
= 2\pi R^2 \gamma n_0 v t
\end{equation}
The step function fixes the limits on the $z$ integration.  The
number of participating nucleons grows linearly with time, and
at time $t_0$ we get $N_i^{\rm part}(t_0) = \pi R^2 L n_0$, which
is the total number of nucleons in the nucleus.

The number of elementary nucleon-nucleon collisions can now be
expressed as
\begin{eqnarray}
N_{1+2 \rightarrow X} &=& \pi R^2 \gamma^2 n_0^2 \int_0^{t_0} dt
\int_{-vt}^{vt} dz \int_0^1 \frac{dx_1}{x_1} \int_0^1 \frac{dx_2}{x_2}
\, v_{12} \, \sigma_{1+2 \rightarrow X}(s_{12}) \nonumber \\
&\,& \sum_{N_1 = 0}^{\infty} \sum_{N_2 = 0}^{\infty} H(x_1,N_1)\,S(N_1,d_1)\,
 H(x_2,N_2)\,S(N_2,d_2) \, .
\end{eqnarray}
Since the nucleons' velocities are antiparallel the velocity factor is
\begin{equation}
v_{12} = \frac{x_1 p_0}{\sqrt{x_1^2 p_0^2 + m_N^2}}
+ \frac{x_2 p_0}{\sqrt{x_2^2 p_0^2 + m_N^2}} \, ,
\end{equation}
where $p_0$ is the beam momentum in the center-of-momentum frame.

The survival function $S(N,d)$ is characterized by the mean free
path $\lambda$ of nucleons in the hot and dense hadronic matter.
For a dilute gas the inverse of the mean free path is the sum of
products of the cross section of the nucleon with the density
of objects it can collide with.
\begin{equation}
\lambda^{-1} = \sum_i \, n_i \sigma_i
\end{equation}
Average particle densities, including baryons and mesons,
were computed in ref. \cite{us} for the hot and dense matter
under consideration.  A plot of the density as a function of
beam energy is shown in Figure 2.  Assuming an average hadron-nucleon
cross section of 25 mb, we find $\lambda$ = 0.4 fm at a laboratory
beam energy of 11.6 GeV/nucleon.  This is very short, and just
emphasizes the physics we discussed in the introduction concerning
hadronic matter versus quark-gluon plasma.

We assume that the collisions suffered by the nucleons are independent
and can be characterized by a Poisson distribution.
\begin{equation}
S(N,d) = \frac{1}{N!} \left( \frac{d}{\lambda}\right)^N
\exp{\left(-\frac{d}{\lambda}\right)}
\end{equation}
Here $d/\lambda$ is the average number of scatterings in a
distance $d$.

The invariant distribution function $H(x,N)$ describes the momentum degradation
of a nucleon propagating through the hot zone. This distribution
function was introduced in the evolution model of Hwa \cite{hwa}.
In this model the nucleon propagates on a straight line trajectory and
interacts with target particles contained within a tube with area
given by the elementary nucleon--nucleon cross section $\sigma_{NN}$.
Csernai and Kapusta \cite{evol} solved the resulting evolution
equations and found that the invariant distribution function in this model
is given by
\begin{eqnarray}
H(x,N) = x \sum_{n=1}^N \left( \begin{array}{c} N\\n \end{array} \right)
w^n (1-w)^{N-n}\,\, \frac{(-\ln{x})^{n-1}}{(n-1)!} +(1-w)^N \, \delta (x-1)
\, .
\end{eqnarray}
The $\delta$-function represents elastic and soft inelastic contributions
to the evolution of the nucleon through the matter.
The probability $w$ is the ratio of inelastic to total nucleon--
nucleon cross section. It corresponds to the probability that the nucleon
scatters inelastically and therefore drops out of the evolution described
by $H$; it is approximately 0.8 in free space. Csernai and Kapusta
found that it reduces to about 0.5 for nucleons propagating through a
nucleus. This value allowed them to obtain a good representation of data with
beam energies in the range of 6-405 GeV. In our case the nucleon is
propagating through hot and dense hadronic matter. We keep $w$ as a free
parameter since we don't know how the value of $w$ changes due to the thermal
excitations and the increased density.

We are interested in the number of pion-producing nucleon-nucleon collisions
with a relatively high center-of-momentum energy squared $s$.  Our
basic result from this section is
\begin{eqnarray}
\frac{dN^{\rm hard}_{\rm in}}{ds} &=& \pi R^2 \gamma^2 n_0^2 \,
\sigma_{\rm in}(s)
\int_0^{t_0} dt \int_{-vt}^{vt} dz
\int_0^1 \frac{dx_1}{x_1} \int_0^1 \frac{dx_2}{x_2} \,
v_{12} \, \delta(s-s_{12}) \nonumber \\
&\,& \sum_{N_1 = 0}^{\infty} \sum_{N_2 = 0}^{\infty} H(x_1,N_1)\,
S(N_1,d_1(z,t))\,H(x_2,N_2)\,S(N_2,d_2(z,t)) \, .
\label{dncoll}
\end{eqnarray}
Here $\sigma_{\rm in}$ is the inelastic nucleon-nucleon cross section,
and $\sqrt{s_{12}}$ is the total energy in the nucleon-nucleon collision
where the nucleons have momentum fractions $x_1$ and $x_2$.

\section{Meson Production Cross Sections}

A phase transition to quark--gluon
plasma will become thermodynamically favorable if the
energy density is large enough. The corresponding phase
boundary in the temperature/chemical potential plane was explored in
\cite{us}. Until now we have only selected nucleon-nucleon scatterings
in which the total available energy $\sqrt{s}$ is large.
In addition, we need to specify what fraction of this energy
goes into meson production.
In this section we estimate the pion number distribution function
$P_n (s)$, which is the probability of producing $n$ pions in a
nucleon--nucleon collision in free space. The pion number
distribution function is linked to the cross section $\sigma_n$
for producing $n$ pions by
\begin{equation}
P_n (s) = \sigma_n (s)/ \sigma_{\rm in}(s) \,.
\label{topo}
\end{equation}
Given $P_n (s)$ we can estimate the number of nucleon-nucleon collisions
that would lead to the production of $n$ pions as
\begin{equation}
N_n = \int_{s_{\rm min}}^{4E_0^2} ds\,P_n (s) \,
\frac{dN_{\rm in}^{\rm hard}}{ds}\,.
\label{number}
\end{equation}
The lower limit of integration is fixed by kinematics and
the upper limit is determined by the beam energy.

We shall approximate the pion number distribution function
$P_n(s)$ with a binomial \cite{comment} and
choose the parameters of this binomial such that we have some rough
agreement with experiment \cite{topo}.
\begin{equation}
P_n (s) = \left( \begin{array}{c} n_{\rm max}\\n \end{array} \right)
\xi^n
(1-\xi)^{n_{\rm max}-n}
\label{bino}
\end{equation}
The maximum number of pions produced in a nucleon-nucleon collision
is determined by kinematics.
\begin{equation}
n_{\rm max} (s)= {\rm Integer} \left(\frac{\sqrt{s}-2m_N)}{m_{\pi}}\right)
\label{nc}
\end{equation}
The parameter $\xi$ is related to the mean multiplicity by
\begin{equation}
\xi(s) = \frac{\langle n \rangle}{n_{\rm max}}
= \frac{3}{n_{\rm max}}\,\left(\frac{1}{4} \langle n_{pp}^- \rangle
+\frac{1}{2} \langle n_{pn}^- \rangle
+\frac{1}{4} \langle n_{nn}^- \rangle\right)\, .
\label{q}
\end{equation}
Here $\langle n \rangle$ is the average pion multiplicity averaged over
$pp$, $pn$ and $nn$ collisions while $\langle n_{pp}^- \rangle$,
$\langle n_{pn}^- \rangle$ and $\langle n_{nn}^- \rangle$ represent
the average negative pion multiplicity in those collisions.
All average multiplicities are functions of $s$, of course. The
factor of 3 is due to isospin averaging.

Experimental data were compiled and parametrized in \cite{multi} as
\begin{eqnarray}
\langle n_{pp}^- \rangle &=& -0.41 + 0.79 F(s) \nonumber \\
\langle n_{pn}^- \rangle &=& -0.14 + 0.81 F(s) \nonumber \\
\langle n_{nn}^- \rangle &=& +0.35 + 0.77 F(s) \, .
\label{paras}
\end{eqnarray}
The function $F$ was introduced by Fermi \cite{fermi},
\begin{equation}
F(s) = \frac{(\sqrt{s}-2m_N)^{3/4}}{s^{1/8}}\, ,
\label{ferf}
\end{equation}
with $s$ measured in GeV$^2$.
The parametrizations in (\ref{paras}) describe the data rather well
except in the threshold region.  We approximate the inelastic
nucleon--nucleon cross section $\sigma_{\rm in}$ by
the inelastic proton--proton cross section. A convenient parametrization
is given in \cite{topo},
\begin{equation}
\sigma_{\rm in} = 30.9 - 28.9\,p_L^{-2.46} - 0.835\,\ln{p_L}
+0.192\,\ln^2{p_L}\, ,
\label{sin}
\end{equation}
where $p_L$ is the laboratory momentum in GeV/c and the cross section
is in mb.  This parametrization is good for $p_L > 0.968$ GeV/c.

The pion production cross sections, as described above, are displayed in
Figure 3.  They have the right shapes and the right orders of magnitude
compared to data \cite{topo}.  However, direct comparison is not possible.
First of all, data generally does not exist for final states with
$\pi^+$, $\pi^-$, and $\pi^0$.  Usually, exclusive experiments can
only measure charged mesons or neutral mesons, not both.  Secondly,
we have not been so sophisticated as to include vector mesons,
the $\eta$ meson, and kaons.  For our purpose such sophistication is
probably not necessary.  We care only about the probability that
a nucleon-nucleon collision leads to a significant amount of energy
release in the sense of conversion of initial kinetic energy to
meson mass.  We are essentially basing our results on the total
inelastic cross section, the average meson multiplicity, kinematics,
and entropy.  Our analysis would be better if we had a handle
on the width of the multiplicity distribution, averaged over the initial
state isospin and summed over the final state isospin.

\section{Star Burst Probabilities}

In this section we put together the ingredients developed in the
last two and compute the number of star bursts which may become
nucleation sites or seeds for plasma formation and growth.

The nucleon-nucleon collisions may be referred to as primary-primary,
primary-secondary, and secondary-secondary, depending on whether
the nucleons have scattered from thermalized particles in the
hot zone (secondary) or not (primary).  The easiest contribution to
obtain is the primary-primary.  All integrations and summations can
be done analytically with the result
\begin{equation}
\frac{dN^{\rm prim-prim}_{\rm in}}{ds} =
4\pi R^2 \sigma_{\rm in}(s) \left(\frac{\lambda \gamma n_0}
{w}\right)^2 \, \left[1-\left(1+w\,\frac{vt_0}{\lambda}\right)\,
\exp{\left(-\frac{vt_0}{\lambda}\right)}\right] \,
\delta (s-4E_0^2) \, .
\end{equation}
The formulas for the primary-secondary and secondary-secondary
contributions can be simplified to some extent but in the end
some summations remain which must be done numerically.

The number of nucleon-nucleon collisions as a function of $s$
are plotted in Figure 4.  Both $w$ = 0.5 and 0.8 are shown;
there is little difference.  The laboratory beam energy is 11.6
GeV per nucleon and the nuclei are gold.  The spike represents
the delta function from primary-primary collisions.  The contribution
from primary-secondary collisions falls from about 11 to 7 GeV$^{-2}$
as $s$ goes from 9 to 25 GeV.  The contribution from secondary-secondary
collisions is almost negligible.

The pion multiplicity distribution arising from these hard collisions
is shown in Figure 5.  It drops by more than nine orders of magnitude
in going from 6 pion production to 18 pion production.  Typically there
is only one hard nucleon-nucleon collision leading to the production
of seven pions in a central gold-gold collision at this energy.

We are interested in the possibility that one of these star bursts
nucleates quark-gluon plasma.  The precise criterion for this to
happen is not known.  However, we can make some reasonable estimates.
In \cite{us} we estimated that a critical size plasma droplet at
these temperatures and baryon densities would have a mass of about
4 GeV.  Any local fluctuation more massive than this would grow
rapidly, converting the surrounding superheated hadronic matter to
quark-gluon plasma.  A similar estimate, based on the MIT bag model,
a simpler hadronic equation of state (free pion gas),
and with zero baryon density,
was obtained much earlier \cite{old}.  Another estimate is obtained
by the argument that at these relatively modest beam energies most
meson production occurs through the formation and decay of baryon
resonances: $\Delta$, $N^*$, etc.  The most massive observed resonances
are in the range of 2 to 2.5 GeV.  Putting two of these in close
physical proximity leads to a mass of 4 to 5 GeV.  We now need an
estimate of the number of pions this critical mass corresponds to.
Let us assume that each particle, nucleon and meson, carries away
a kinetic energy equal to one half its rest mass.  If a particle
would have too great a kinetic energy then it might escape from
the nucleon-nucleon collision volume long before its neighbors and
so would not be counted in the rest mass of the local fluctuation.
Taking 4 GeV, dividing by 1.5, and subtracting twice the nucleon mass
leaves about 6 pion rest masses.  So our most optimistic estimate
is that one needs a nucleon-nucleon collision which would have led
to 6 pions if it had occurred in free space.  One might be less
optimistic and require the production of 8 or 10 pions instead.

In Figure 6 we show the total number $N_>$ of nucleon-nucleon collisions
which would lead to the production of at least $n_{\rm crit}$ pions.
We may view $n_{\rm crit}$ as the minimum number necessary to form
a nucleation site or plasma seed.  If $n_{\rm crit}$ = 6 is the
relevant number then there are on average 7 such nucleon-nucleon
collisions per central gold-gold collision.  If 8 or 10 are the
relevant multiplicities then there is only one such critical star
burst every 1 or every 25 central gold-gold collisions, respectively.
These numbers vary somewhat with $w$; the numbers quoted are averages.
Conservatively, we may conclude that the probability of at least
one plasma seed appearing via this mechanism is in the range of
1 to 100\% per central gold-gold collision at the highest energy
attainable at the AGS.  These probabilities are about one to two
orders of magnitude greater than those estimated in \cite{us}
on the basis of thermal homogeneous nucleation theory.

\section{Consequences for the Multiplicity Distribution}

The results of the last section confirm the possibility of producing
quark--gluon plasma droplets in rare events at AGS. Once formed the droplets
grow rapidly due to the significant superheating of the hadronic matter. This
process was explored in~\cite{us} where it was found that the radii of such
droplets can reach $3-5$ fm.  Since the phase transition
is occurring so far out of equilibrium we would expect a significant
increase in the entropy of the final state.  This could be seen in the
ratio of pions to baryons, for example, or in the ratio of deuterons
to protons \cite{me}.  Along with the increased entropy should come
a slowing down of the radial expansion due to a softening in the
matter, that is, a reduction in pressure for the same energy
density.  Together, these would imply a larger source size and
a longer lifetime as seen by hadron interferometry \cite{scott}.

In this section we study one of the experimental ramifications in detail.
Specifically, we look at the charged particle multiplicity distributions and
investigate under what conditions one might be able to detect the rare events
from the structure of this distribution.

In Figure 7 we plot the ratio of entropy to total baryon number
$S/B$ for the hadronic and quark--gluon plasma phase for fixed beam
energies. Fixed beam energy means that initially both the energy density
and the baryon number density of the system is given which then determine
the corresponding entropies via the equation of state. We use the equation
of state discussed in~\cite{us} for all further calculations.
It is helpful to consider two extreme and opposite scenarios. Either the
matter stays all the time in the hadronic phase, or the matter has
been completely converted to quark--gluon plasma by the time $t_0$ and
only hadronizes later.
The difference of the entropies produced in these two scenarios is given by the
difference of the two curves in Figure 7. It represents an upper limit on the
additional number of pions produced. Since the temperature is comparable to
or larger than the pion mass the excess entropy is proportional to the
maximum number of excess pions
\begin{eqnarray}
3 \frac{\Delta N_{\rm -}}{B} = \frac{1}{3.6} \frac{\Delta S}{B}\, .
\label{npi}
\end{eqnarray}
The number of additional negatively charged pions per baryon
$\Delta N_{\rm -}/B$ is linearly related to the entropy difference $\Delta S$
determined from Figure 7.
The result is shown in Figure 8 for central Au + Au collisions. At beam
energies of $11.6$ GeV/A we produce $0.33$ additional negatively charged pions
per participating baryon.  This is an upper limit, and in reality we
would expect less.

These additional mesons might be visible in the charged
particle multiplicity distribution which would have the form
\begin{eqnarray}
P_n = (1-q) \, P_{\; n}^{\rm had} (N_{\rm had})
	+ q \, P^{\rm qg}_{\; n} (N_{\rm qg})\, .
\label{double}
\end{eqnarray}
Here $q$ is the probability of finding a central event in which plasma
is formed, $P_{\; n}^{\rm had}$ is the multiplicity distribution
for purely hadronic events with mean $N_{\rm had}$, and $P^{\rm qg}_{\; n}$
is the multiplicity distribution for events in which a plasma
was formed with mean $N_{\rm qg}$.

Experimentally one would expect to see a bump in $P_n$ at larger values
of $n$. A structure like that was found in charged particle multiplicity
distributions in $p\bar{p}$ collisions at the CERN \cite{ua5,fuglesang} and
Fermilab \cite{cdf,e735} colliders. For energies larger then 540 GeV a
shoulder develops in the multiplicity distribution, becoming more pronounced
as the beam energy increases. It is assumed that this structure is due
to the onset of minijets. It is definitely an indication of new physics.

In Figure 9 we plot the charged particle multiplicity distribution for
$p\bar{p}$ collisions at $\sqrt{s} = 900$ GeV from the UA5
collaboration~\cite{ua5}.
For energies less then 500 GeV it was found that the distribution could be well
described by a negative binomial distribution of the form
\begin{eqnarray}
P_n (\bar{n}, k) =
\left( \begin{array}{c} n+k-1\\ k-1 \end{array} \right)
\left[ \frac{ \bar{n} / k }{ 1+(\bar{n} / k)} \right]^n
\frac{1}{[1+(\bar{n} / k)]^k} \, .
\label{nbd}
\end{eqnarray}
The parameter $k$ characterizes the width of the distribution. For $k
\rightarrow \infty$ we recover a Poisson distribution, the distribution with
the smallest width.  One can see from the figure that at 900 GeV a single
negative binomial (NBD) cannot describe the data anymore.
A double negative binomial (DNBD) of the form
discussed in eq. (\ref{double}) on the other hand describes it very well.
The question remains to what extent a similar analysis might be able to reveal
rare events of quark--gluon plasma production at AGS.

A rough criteria for the observability of such structure in distributions of
the form (\ref{double}) is
\begin{eqnarray}
\frac{2}{\sqrt{N_{\rm bin}}} P_{N_{\rm qg}}^{\rm had} =
q P_{N_{\rm qg}}^{\rm qg} \, .
\label{criteria}
\end{eqnarray}
Here $N_{\rm bin}$ is the number of observed central Au + Au collisions
for which the central multiplicity of the bin is $N_{\rm qg}$.
The right--hand side of eq.
(\ref{criteria}) is the magnitude of the rare events to the overall
multiplicity, while the left hand side gives the statistical resolution.
The assumption here is that $q$ is small, so that at $N_{\rm qg}$ we
can use $P_n \sim P_n^{\rm had}$ for the left--hand side.

To obtain a feeling for the shape and applicability of eqs. (\ref{nbd})
and (\ref{criteria}) we plot in Figures 10 and 11 different negatively
charged particle multiplicity distributions as might be expected for
central Au + Au collisions at AGS with $E_{\rm beam} = 11.6$ GeV/A. From
{}~\cite{multi} we obtained the mean for purely hadronic events to be
$N_{\rm had} = 145$. This is slightly larger than the value
$N_{\rm had} = 131 \pm 21$ cited in \cite{multi} for $355 \pm 7$
participating nucleons since we are assuming that all $2 A$ nucleons
are participating in the collision. The result depicted in Figure 8
for the upper limit on the additional number of negatively charged
pions produced per participating baryon allows us to deduce an upper
limit of $N_{\rm qg} = 193$ on the mean for the events with quark--gluon
plasma production. In Figure 10 we plot the
negatively charged particle multiplicity distribution defined in eq.
(\ref{double}) for different values of the probability $q$. We use Poisson
distributions for $P^{\rm had}$ and $P^{\rm qg}$ and take the upper limit
for rare events $N_{\rm qg} = 193$ as the mean for $P^{\rm qg}$.
A shoulder develops for small $q$ and becomes more pronounced the
larger $q$ is. In Figure 11 we fix $q=0.1$ and investigate the effect
of different values of the mean $N_{\rm qg}$ of the distribution for events
with some quark--gluon plasma production. If this mean is close to the mean
of purely hadronic events we will only find some broadening of the overall
distribution. This would be the case if the phase transition is weakly
first order or second order.  For larger $N_{\rm qg}$ we begin to see
a well established shoulder develop. For large $N_{\rm qg}$ a second
maximum appears.

It is clear that the exact values of the probability $q$ and of
the mean $N_{\rm qg}$ of rare events will
be crucial for the experimental observation of a phase transition.
We have provided a first glimpse into this problem, but in the end
it is up to experiment to discover
new physics in multiplicity distributions at the AGS.

\section{Summary and Conclusion}

We have estimated the probability that hard nucleon-nucleon collisions
initiate the formation of seeds of quark-gluon plasma at AGS energies.
Based on our previous studies we know that these will grow rapidly
to convert most of the superheated hadronic matter to quark-gluon plasma.
Our estimates are based on reasonable assumptions and approximations
to the kinetic theory of hadronic physics.  Better estimates could
be made using event simulators like RQMD and ARC together with
more detailed knowledge of multi-particle production in nucleon-nucleon
collisions.  We find that anywhere from 1\% to 100\% of central
Au + Au collisions should lead to significant quark-gluon plasma
formation.  A major assumption is that there is a phase transition
and that it is first order.

We have already proposed that the formation of plasma in rare events
should have an observable consequence for hadron interferometry,
deuteron production, and the meson multiplicity distribution.
In this paper we have studied the effect on the multiplicity
distribution.  It would be observable as a shoulder or second
maximum at some multiplicity higher than the most probable one.
If there is a phase transition but it is second order or weakly
first order then the effect will be much more difficult to see.
We eagerly await the results of experiments.

\section*{Acknowledgements}

We thank R. Venugopalan and C. J. Waddington for stimulating discussions
and L. Csernai and P. Lichard for comments on the manuscript.
This work was supported by the U.S. Department of Energy under
grant number DE-FG02-87ER40328.


\newpage

\section*{Figure Captions}

Figure 1: Schematic of a central collision between two nuclei.
Nucleus 1 is incident from the left and nucleus 2 is incident from
the right.  The shaded area is the hot and dense overlap zone
which is expanding along the beam axis with the original beam
velocity $v$.  A hard nucleon-nucleon collision leading to a
large energy release, or star burst, is indicated at the longitudinal
position $z$.\\

\noindent Figure 2: The density of hadrons $n_{\rm tot}$ in the hot and dense
overlap zone as a function of laboratory beam energy $E_{\rm beam}$.\\

\noindent Figure 3: The pion production cross sections $\sigma_{\rm n}$
versus energy $s$ as computed according to the text.  Note that they are
averaged over initial state isospin.\\

\noindent Figure 4: Distribution $dN/ds$ in $s$ of hard nucleon-nucleon
collisions taking place in the hot zone.  The value of $w$
is 0.5 (4a) and 0.8 (4b).\\

\noindent Figure 5: Number of hard nucleon-nucleon collisions $N_n$
leading to a particular final state pion multiplicity $n$.\\

\noindent Figure 6: Number of hard nucleon-nucleon collisions $N_{\rm >}$
with at least $n_{\rm crit}$ pions produced.\\

\noindent Figure 7: Ratio of entropy to baryon number $S/B$ for
fixed beam energy.\\

\noindent Figure 8: Upper limit on the additional number of negative pions
produced per participating baryon $\Delta N_{\rm -}/B$ in central
Au + Au collisions as a function of beam energy.\\

\noindent Figure 9: Charged particle multiplicity distribution for
$\bar{p} p$ collisions at $\sqrt{s}=$ 900 GeV. The parameters for
the fits are taken from \cite{fuglesang} with a probability of the
second, high multiplicity, component being 0.35.\\

\noindent Figure 10: Negatively charged particle multiplicity distribution
for central Au+ Au collisions at $E_{\rm beam} = 11.6$ GeV/A for
different values
of the probability q. The mean for purely hadronic events is taken to be
$N_{\rm had} = 145$ while the mean for events with quark--gluon plasma
production is taken to be $N_{\rm qg} = 193$.\\

\newpage

\noindent Figure 11: Negatively charged particle multiplicity distribution
for central Au + Au collisions at $E_{\rm beam} = 11.6$ GeV/A for different
values of the mean $N_{\rm qg}$ for rare events.  The probability
is fixed at $q=0.1$ and the hadronic mean multiplicity is fixed
at $N_{\rm had} = 145$.\\


\begin{thebibliography}{99}

\bibitem{rqmd} H. Sorge, H. St\"ocker and W. Greiner, Annals Phys.
(NY) {\bf 192}, 266 (1989); R. Mattiello, H. Sorge, H. St\"ocker
and W. Greiner, Phys. Rev. Lett. {\bf 63}, 1459 (1989);
H. Sorge, A. V. Keitz, R. Mattiello, H. St\"ocker and W. Greiner,
Phys. Lett. {\bf 243B}, 7 (1990); A. V. Kreitz, L. Winckelmann,
A. Jahns, H. Sorge, H. St\"ocker and W. Greiner, Phys. Lett.
{\bf 263B}, 353 (1991).

\bibitem{arc} Y. Pang, T. J. Schlagel and S. H. Kahana, Phys. Rev.
Lett. {\bf 68}, 2743 (1992); T. J. Schlagel, Y. Pang and S. H.
Kahana, Phys. Rev. Lett. {\bf 69}, 3290 (1992); S. H. Kahana,
Y. Pang, T. J. Schlagel and C. Dover, Phys. Rev. C {\bf 47},
R1356 (1993).

\bibitem{pang} Y. Pang, proceedings of {\it Quark Matter '95}
(Monterey, CA, Jan. 1995) to appear.

\bibitem{us} J. I. Kapusta, A. P. Vischer and R. Venugopalan, Phys. Rev.
C {\bf 51}, 901 (1995).

\bibitem{hwa} R. C. Hwa, Phys. Rev. Lett. {\bf 52}, 492 (1984).

\bibitem{evol} L. P. Csernai and J. I. Kapusta Phys. Rev. D {\bf 29},
2664 (1984); ibid. {\bf 31}, 2795 (1985).

\bibitem{comment} There is always some uncertainty in how much of
the $n$ = 0 term in the binomial is to be associated with diffractive
and elastic events; this affects the exact measure of $\sigma_{\rm in}$
as used in eq. (15).  These uncertainties are relatively minor for us.

\bibitem{topo} A. Baldini, V. Flaminio, W. G. Moorhead and
D. R. O. Morrison, in {\it Landolt--B\"{o}rnstein},
New Series I/12b, Springer, 1988, pp. 149--180.

\bibitem{multi} M. Ga\'{z}dzicki and D. R\"{o}hrich, Z. Phys.
C {\bf 65}, 215 (1995).

\bibitem{fermi} E. Fermi, Prog. Theor. Phys. {\bf 5}, 570 (1950).

\bibitem{old} J. I. Kapusta, Phys. Lett. {\bf 143B}, 233 (1984).

\bibitem{me} P. J. Siemens and J. I. Kapusta, Phys. Rev. Lett.
{\bf 43}, 1486 (1979).

\bibitem{scott} S. Pratt, Phys. Rev. D {\bf 33}, 1314 (1986).

\bibitem{ua5} UA5 Collaboration, Z. Phys. C {\bf43}, 357 (1989).

\bibitem{fuglesang} C. Fuglesang, for the UA5 Collaboration:
``UA5 Multiplicity Distributions and Fits of Various Functions",
proceedings of {\it Multiparticle Dynamics: Festschrift
for Leon van Hove} (La Thuile, Valle d'Aosta, Italy, March 1989)
eds. A Giovannini and W. Kittel, World Scientific, 1990.

\bibitem{cdf} F. Rimondi, for the CDF Collaboration: ``Multiplicity
Distributions in \={p}p Interactions at $\sqrt{s} =$ 1800 GeV",
proceedings of the {\it XXIII International Symposium on
Multiparticle Dynamics} (Aspen, CO, Sept. 1993) eds. M. M. Block
and A. R. White, World Scientific, 1994.

\bibitem{e735} C. S. Lindsey, for the E735 Collaboration, Nuc.
Phys. {\bf A544}, 343c (1992).

\end{thebibliography}
\end{document}